# Towards reducing tension-compression yield and cyclic asymmetry in pure magnesium and magnesium-aluminum alloy with cerium addition

Shubham Sisodia, Jananandhan S, Vamsi Krishna Pakki, Chethan Konkati, Ankur Chauhan*

Department of Materials Engineering, Indian Institute of Science, Bengaluru, Karnataka-560012, India

*Corresponding author: ankurchauhan@iisc.ac.in

**Abstract**

In this study, we report the effect of cerium (Ce) addition on the tension-compression yield and cyclic asymmetry in commercially pure magnesium (Cp-Mg) and Mg-Al alloy at room temperature (RT). The investigated materials Cp-Mg, Mg-0.5Ce, and Mg-3Al-0.5Ce were extruded at 400°C, followed by annealing at the same temperature for one hour. Incorporating 0.5wt.% Ce in pure Mg results in the weakening of its basal texture, uniform distribution of $Mg_{12}Ce$ precipitates, and grain size refinement. Consequently, the tensile yield strength and ductility of pure Mg increased, and tension-compression yield asymmetry was eliminated. However, the presence of 3wt.% Al in Mg suppresses the beneficial effects of Ce addition. The formation of complex precipitates, such as Mg-Al-Ce and $Al_{11}Ce_3$, limits the weakening of the basal texture, reduction in grain size, improvement in ductility, and elimination of tension-compression yield asymmetry observed in Mg-0.5Ce. Nevertheless, Al contributes to the solid solution strengthening in Mg and possibly lowers the critical stress required for twinning in Mg, resulting in the highest tensile strength of Mg-3Al-0.5Ce. Finally, the addition of 0.5wt.% Ce enhances the cyclic strength, stabilizes cyclic stress response, reduces inelastic strain, and minimizes cyclic asymmetry in both pure Mg and Mg-Al alloy while maintaining a comparable fatigue life. Overall, Ce addition positively impacts the microstructure and mechanical behavior of pure Mg and its investigated alloy. The reasons for these improvements are discussed in detail.

## 1. Introduction

Magnesium (Mg), a lightweight metal, can potentially reduce the weight of automobile and aerospace components [1]. However, compared to conventional alloys such as steel and aluminum alloys, wrought Mg alloys have poor ductility and are less formable at room temperature (RT) due to their large anisotropy and lack of sufficient slip systems. The number of operative slip systems determines the ductility in any alloy system. Mg having hexagonal close-packed (HCP) lattice, in general, can deform by slip on basal (0001) <11$\bar{2}$0>, prismatic {10$\bar{1}$0} <11$\bar{2}$0>, and pyramidal I {10$\bar{1}$1} <10$\bar{1}$2> or pyramidal II {$\bar{1}\bar{1}$22} <11$\bar{2}$3> (c+a) slip systems or by twinning. Twinning occurs either through the formation of extension {10$\bar{1}$2} <10$\bar{1}$1> or contraction twins {10$\bar{1}$1} <10$\bar{1}$2>, depending on the strain path, and can accommodate strain along the c-axis. For instance, extension twins are formed when the tensile strain component is parallel to the c-axis, and contraction twins are formed when the compressive strain component is parallel to the c-axis [2]. In addition, double twinning can also occur, resulting in the development of a {10$\bar{1}$2} extension twin within a pre-existing contraction twin, which causes a contraction along the c-axis. Previous studies have reported the critical resolved shear stress (CRSS) values for slip and twinning systems in Mg and its alloys at RT [3,4]. It is generally accepted that the CRSS values at RT follow the following trend: $CRSS_{basal} < CRSS_{extension\ twinning} < CRSS_{prismatic} < CRSS_{pyramidal} < CRSS_{contraction\ twinning}$. At RT, basal slip and extension twinning are the dominant deformation modes due to their relatively low CRSS values [5–7]. CRSS for basal slip is approximately 1/100 of the non-basal slip systems such as prismatic and pyramidal [8–11]. Consequently, a strong basal texture generally develops in Mg alloys during processing, such as extrusion, rolling, and forging, which can persist even after post-annealing treatments [6]. This gives rise to anisotropy in the mechanical properties, reduced ductility, and limited formability at RT. Therefore, to achieve isotropic properties and sufficient ductility, at least five independent slip systems need to be activated, including non-basal slip systems [12]. This is mainly done by randomizing basal texture.

The two primary methods used to randomize the basal texture of Mg alloys include post-processing treatments or the addition of rare-earth (RE) elements. Post-processing treatments, such as extrusion followed by differential speed rolling (DSR) [13–15] or asymmetric rolling by DSR in AZ31 [16–18], have been shown to weaken the basal texture. Alternatively, another effective method is the alloying of RE elements like Yttrium (Y), Neodymium (Nd), Gadolinium (Gd), Lanthanum (La), Cerium (Ce), etc. The weakening of the basal texture is attributed to changes in the bonding energy between Mg and RE atoms and alterations in the c/a ratio of the magnesium's HCP lattice. These changes are suggested to restrict basal slip and twinning while promoting non-basal slip activity [19,20]. Several studies have demonstrated that adding RE elements in Mg alloys enhances strength and ductility and reduces tension-compression yield asymmetry [21–27].

Additionally, while considerable progress has been made in understanding the monotonic behavior of Mg-RE alloys, there has been a limited focus on investigating their strain-controlled low-cycle fatigue (LCF) response until recently. For instance, Mirza et al. [28–30] extensively studied the strain-controlled LCF behavior of extruded Mg-10Gd-3Y-0.5Zr (GW103K) alloy and showed symmetrical hysteresis loops, suggesting the absence of the tension-compression yield asymmetry. Wang et al. [31,32] explored the LCF behavior of the extruded Mg-8Gd-3Y-0.5Zr (GW83) alloy and showed fatigue crack initiation at the grain boundaries. Mokdad and Chen [33,34] investigated the cyclic deformation characteristics of a low (0.2 wt.%) Nd-containing ZEK100-O alloy, demonstrating reduced hysteresis loop asymmetry and extended fatigue life compared to RE-free Mg alloys. Li et al. [35,36] examined the deformation modes, transferability, and in-plane anisotropy during LCF of Mg and Mg–3Y alloy, highlighting the texture-dependent anisotropy in mechanical properties and the activation of multiple slip systems with Y addition while suppressing twinning-detwinning activity. Overall, these studies collectively show the positive impact of RE element addition on the cyclic deformation behavior of Mg-RE alloys, with the specific behavior influenced by the RE element content and processing route employed.

Among rare-earth elements, Ce has a relatively low solubility in pure magnesium, ~ 0.52 wt.%, at the eutectic temperature of 590°C [37]. Considering that the solubility of Ce in magnesium is even lower at RT, it enables the precipitation of Mg-Ce precipitates alongside the small percentage of Ce in the solid solution. This indicates that by strategically alloying magnesium with an appropriate percentage of Ce makes it feasible to utilize multiple strengthening mechanisms, including solid solution strengthening, precipitation strengthening, as well as Hall-Petch strengthening. These mechanisms are expected to show promising outcomes in modifying the basal texture of magnesium, enhancing its strength and ductility, and improving its cold workability [38–42]. LCF studies on Mg-Ce alloys are rare. Only a few studies have reported on the LCF behavior of Mg-Ce alloys, such as hot rolled Mg-Mn-Ce (ME20M-H112) alloy [43] and Mg-9Al-0.86Zn-0.12Mn (AZ91D) alloy with varying Ce content [44]. Unlike high Gd-based Mg alloys (e.g., GW103K), the investigated Mg-Ce alloys exhibited asymmetrical hysteresis loops like RE-free extruded Mg alloys. This asymmetry is mainly attributed to twinning-detwinning activities during cyclic deformation, which warrant further optimization of the Mg-Ce alloys microstructure and composition. To the best of the authors' knowledge, no comprehensive study is available that investigates the overall effect of 0.5wt.% Ce addition on the microstructure, tension-compression behavior, and the LCF response of commercially pure Mg (Cp-Mg) and Mg-3wt.%Al alloy. Consequently, this study investigates the microstructure, tensile, compression, and strain-controlled LCF responses of Cp-Mg, Mg-0.5Ce, and Mg-3Al-0.5Ce alloys.

## 2. Experimental

### 2.1 Materials and mechanical testing

The investigated Cp-Mg and Mg-0.5Ce, and Mg-3Al-0.5Ce alloys ingots were cast and homogenized at 400°C for 24h. The homogenized ingots of diameter 60 mm were subsequently extruded in the form of bars of cross-section 16 × 16 mm$^2$ at 400°C. Thereafter, flat dog-bone-shaped tensile specimens with a gauge length of 6 mm and thickness of 1 mm and compression specimens with dimensions of 3 × 3 × 4.5 mm$^3$ were prepared along the extrusion direction. For LCF testing, standard M8 cylindrical specimens were also prepared along the extrusion direction with a gauge length of 14 mm and a gauge diameter of 4 mm. All specimens were then vacuum sealed in quartz tubes and heat treated at 400°C for 1h to obtain as recrystallized microstructures. Before testing, all specimens' gauge sections were ground till 4 μm silicon carbide abrasive papers. The tensile and compression tests were conducted on Instron 5967 universal testing machine (UTM). Strain under both tension and compression loading was determined by monitoring the UTM crosshead displacement. LCF tests were carried out on a computer-controlled BISS servo-electric testing machine featuring a 12.5 mm gauge length extensometer. The axial push-pull strain-controlled fatigue tests were conducted in air, employing a symmetrical triangular waveform (with R = -1) and a strain amplitude of ± 0.5%. All tests were performed at RT under a nominal strain rate of 10$^{-3}$ s$^{-1}$. The elastic modulus determined from the LCF tests was applied to adjust the monotonic stress-strain curves. Tests at each condition were repeated three times to get a sense of experimental scatter.

### 2.2 Microstructural characterization

The specimens were prepared along the extrusion direction for microstructural characterization of both as recrystallized and deformed conditions. The specimens were initially ground till 4 μm silicon carbide abrasive papers followed by polishing with colloidal silica suspension of particle size of 0.06 μm. Thereafter, specimens were electropolished at 3V for 30s and at 1.5V for 2min. The electrolyte contained three parts H$_3$PO$_4$ and five parts ethanol, which was cooled to within 5 to 10°C.

For electron backscattered diffraction (EBSD) investigations, a Carl Zeiss Gemini 450 field emission-based scanning electron microscope (FESEM) equipped with an EDAX detector was employed. EBSD data was acquired at an accelerating voltage of 25 kV, 6 nA probe current, and a step size of 1.3 μm. EBSD assisted in characterizing texture, grain size, and deformation-induced slip and twinning mechanisms. The investigation of slip activity involving the activation of basal and/or non-basal slip systems was conducted through slip trace analysis [45–49]. In this experimental approach, the surface of the sample subjected to tensile testing was imaged to identify slip traces, defined as straight lines resulting from the interaction of active slip planes with the sample's surface. Subsequently, the orientation of grains containing slip traces was determined using EBSD. Thereafter, each trace orientation was compared with potential traces corresponding to various slip systems. These potential traces were generated using a dedicated MATLAB code [50], which required the input of the grain's orientation in terms of three Euler angles. These angles were obtained from the post-test EBSD inverse pole figure (IPF) map. The MATLAB code then facilitated the selection of the best-matching slip system for a given trace based on orientation comparison. Additionally, the MATLAB code provided information regarding the Schmid Factor associated with the corresponding slip system.

X-ray diffraction (XRD) bulk texture measurements were conducted using a Bruker D8 Advance diffractometer equipped with a high-resolution area detector, operating at 40 kV and 40 mA, using filtered ion radiation and poly capillary focusing optics. The X-ray beam size was 2 mm. The measurements were performed on the specimen mid-plane to obtain the bulk texture and avoid any surface effects. A set of six measured pole figures (i.e., ranging from $\psi=$ 0° to 75°) (0002), (10$\bar{1}$0), (10$\bar{1}$1), (10$\bar{1}$2), (11$\bar{2}$0), and (10$\bar{1}$3) was employed to calculate the orientation distribution function (ODF) using a quantitative texture analysis using a Bunge notation of the Euler angles with the help of (ResMat) Textools software.

The wavelength-dispersive X-ray spectroscopy (WDS) was carried out for precipitates composition analysis using a JEOL JXA 8230 electron probe microprobe analyzer (EPMA) with a tungsten filament source. Measurements were performed with an accelerating voltage of 15 kV, probe current of 50 nA, and peak counting time of 10 sec. Calibration of the instrument was carried out using natural and pure metal standards. The Mg-K$\alpha$, Al-K$\alpha$, and Ce-K$\alpha$ characteristic X-ray peak intensities were measured using pure Mg, Al, and Ce as standards, respectively. The data reduction was performed using the ZAF model correction procedure.

## 3. Results and discussion
### 3.1 Effect of Ce addition on the microstructure

The EBSD IPF maps acquired along the normal direction (ND) of the extruded and annealed Cp-Mg, Mg-0.5Ce, and Mg-3Al-0.5Ce alloys are shown in Fig. 1. Evidently, all materials exhibit fully recrystallized microstructure with distinctly different grain sizes. Cp-Mg displays the largest average grain size of ~ 76 μm followed by ~ 26 μm of Mg-3Al-0.5Ce and ~ 13 μm of Mg-0.5Ce as shown in Fig. 2. Additional analysis of IPF maps also reveals distinctly different textures in the investigated materials. IPF map of Cp-Mg in Fig. 1a clearly shows predominantly red-colored grains suggesting a strong basal texture, i.e., basal planes (0002) are prominently aligned parallel to the extrusion direction. In contrast, IPF maps of Mg-Ce alloys in Fig. 1b and Fig. 1c reveal a weakening of basal texture. Noticeably, grains with colors other than red often appear, particularly in the case of Mg-0.5Ce. This suggests that Mg-0.5Ce exhibits a more substantial basal texture reduction than Mg-3Al-0.5Ce.

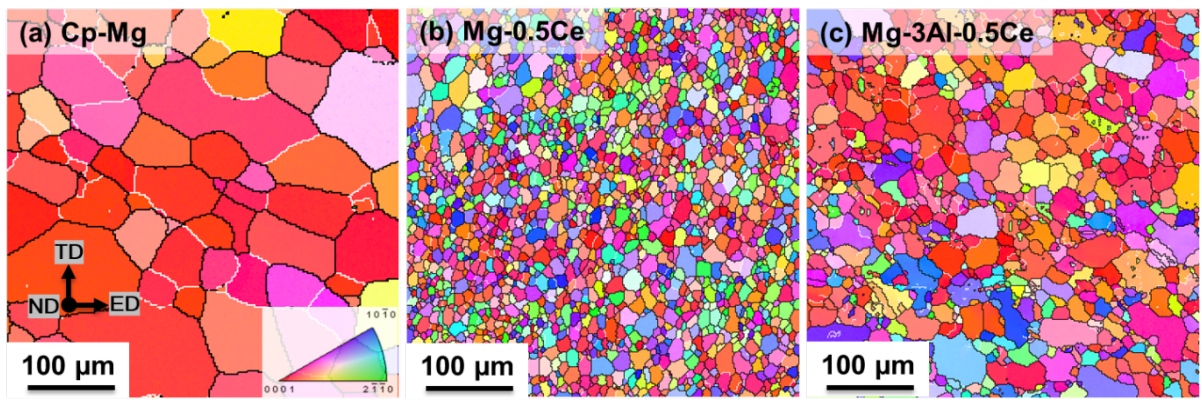

Fig. 1: IPF maps obtained via EBSD along the normal direction (ND) for annealed (a) CP-Mg, (b) Mg-0.5Ce, (c) Mg-3Al-0.5Ce. A common color key is provided as an inset in the lower left corner of the IPF map in (a). An apparent reduction in grain size and basal texture (as grains with colors other than red often appear) is noticeable upon Ce addition. Mg-0.5Ce exhibits a more substantial decrease in basal texture compared to Mg-3Al-0.5Ce.

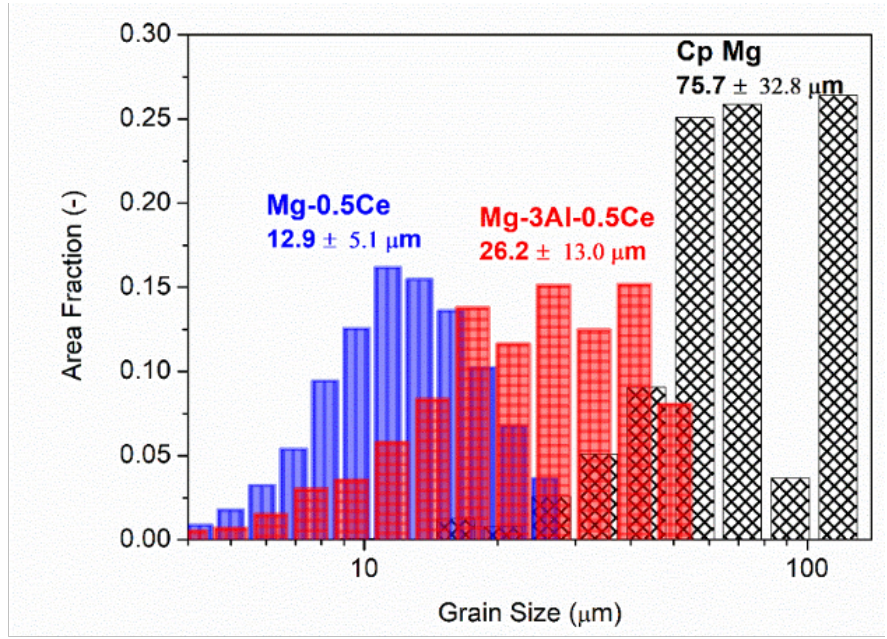

Fig. 2: Bar chart presents grain size distribution of Cp-Mg and its investigated Mg-Ce alloys (Mg-0.5Ce and Mg-3Al-0.5Ce). A clear reduction in grain size distribution is apparent upon Ce addition. Cp-Mg exhibits the largest average grain size of ~ 76 μm followed by ~ 26 μm of Mg-3Al-0.5Ce and ~ 13 μm of Mg-0.5Ce.

XRD was employed for bulk texture investigations to complement and validate the outcomes from EBSD results. Fig. 3 presents the XRD (0002), ($10\bar{1}0$), and ($11\bar{2}0$) pole figures of Cp-Mg and its investigated Mg-Ce alloys obtained along the normal direction. As evident from Fig. 3a, Cp-Mg exhibits a strong intensity around the center of the (0002) basal pole, with a maximum value of 6.9. The intensities along the ($10\bar{1}0$) and ($11\bar{2}0$) poles (Fig. 3b and 3c) are relatively modest and evenly distributed, with maximum values of 1.8 and 1.7, respectively. Compared to the (0002) pole, this observation indicates that nearly all grains c-axes are oriented perpendicular to the extrusion direction. This observation aligns with the typical magnesium texture formed through dynamic recrystallization during extrusion.

For Mg-3Al-0.5Ce, as shown in Fig. 3d, the (0002) basal pole is centered but extended along the TD direction. Moreover, like Cp-Mg, relatively weak and distributed intensities are visible along the ($10\bar{1}0$) and ($11\bar{2}0$) poles (depicted in Fig. 3e and 3f, respectively). Furthermore, the maximum intensity of the basal texture (5.3) is slightly weakened compared to Cp-Mg (6.9), suggesting that in Mg-3Al-0.5Ce, relatively fewer grains have their c-axes perpendicular to the extrusion direction.

In contrast to both Cp-Mg and Mg-3Al-0.5Ce, Mg-0.5Ce showcases a notably diffused and widespread basal texture along the (0002) pole (Fig. 3g). Additionally, the (10$\bar{1}$0) and (11$\bar{2}$0) poles, shown in Fig. 3h and Fig. 3i, respectively, exhibit diffuse patterns with maximum intensities that are comparable to that of the (0002) pole. The basal pole's maximum intensity in Mg-0.5Ce (1.9) is the lowest among Cp-Mg and Mg-3Al-0.5Ce. This highlights the fact that the incorporation of Ce alone results in the most substantial reduction of basal texture within wrought Mg, leading to a relatively more randomized texture. This contrasts with the Mg-3Al-0.5Ce scenario, where the Ce's positive effect is suppressed when introduced alongside substitutional elements such as Al. These bulk texture findings are in accordance with the micro-texture data obtained through EBSD (see Fig. 1).

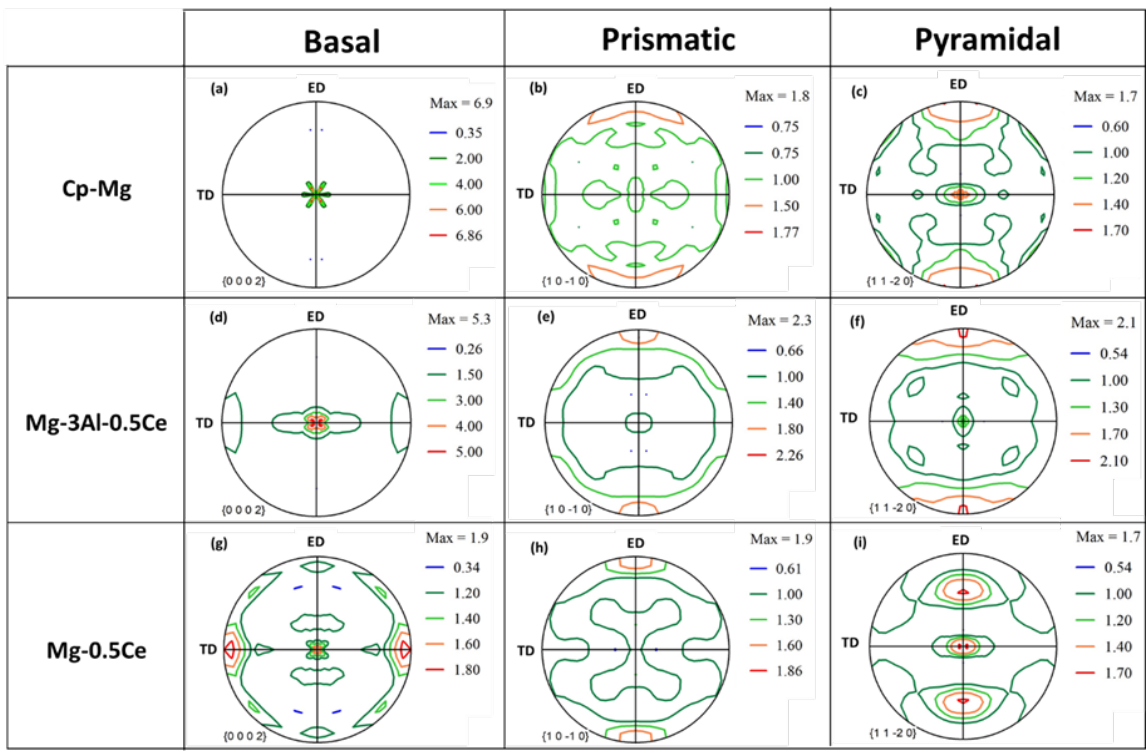

Fig. 3: (0002), (10$\bar{1}$0), and (11$\bar{2}$0) pole figures obtained via XRD along the normal direction (ND) of extruded and annealed (a, b, c) Cp-Mg, (d, e, f) Mg-3Al-0.5Ce, and (g, h, i) Mg-0.5Ce, respectively. A strong intensity at the center of the basal pole for Cp-Mg gets reduced and distributed upon Ce addition. In the presence of Al, the Ce beneficial effect is relatively suppressed as the basal pole gets only spread along the transverse direction (TD).

EPMA investigations were carried out to know the reasons behind the distinct microstructure evolution of two investigated Mg alloys. Fig. 4 and Fig. 5 present EPMA results obtained from Mg-0.5Ce and Mg-3Al-0.5Ce alloys, respectively. As evident from Fig. 4, Mg-0.5Ce contains uniformly distributed precipitates rich in both Mg and Ce. These sub- to few-microns-sized precipitates are present both inside the grains and at the grain boundaries (GBs). Additionally, Ce segregation at GBs is also expected. It is known that the Ce maximum solubility limit in pure Mg at eutectic temperature (590°C) is 0.52 wt.% [37], which will decrease with a decrease in temperature. Therefore, it can be concluded that the Ce content in the present alloy (0.5 wt.%) exceeds its room temperature solubility limit in pure Mg, resulting in the precipitation of Mg-Ce precipitates during solidification. Quantitative analysis of the EMPA results reveals that these precipitates are mainly the $Mg_{12}Ce$ phase, which has a body-centered tetragonal structure. It is known that such precipitates cannot deform compatibly with the matrix during extrusion, leading to stress concentration near precipitates [51]. The subsequent plastic relaxation relieves the stress build-up around the precipitates, which assist in nucleating strain-free grains during recrystallization with relatively random orientation [51,52]. This, therefore, weakens the basal texture as seen from our micro- and bulk-texture analysis (see Fig. 1b and Fig. 3g). Additionally, Mg-Ce-rich precipitates can pin GBs, which inhibit grain growth, leading to Mg-0.5Ce finer grain size in comparison to Cp-Mg [53]. Similar observations with the weakening of basal texture and grain refinement by RE elements addition [51,54], like yttrium [55], neodymium [56], gadolinium [57], and erbium [58], have been reported before.

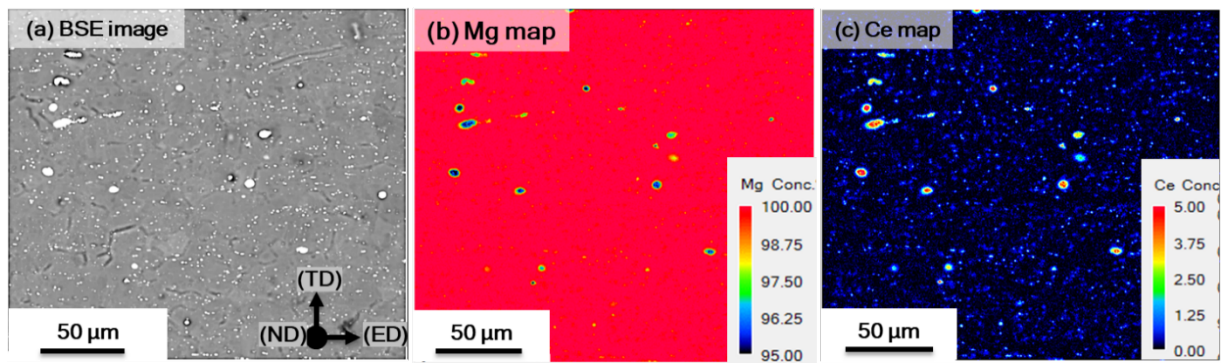

Fig. 4: EPMA results for Mg-0.5Ce alloy: (a) Backscattered electron (BSE) micrograph, (b) elemental Mg map, and (c) elemental Ce map. Bright contrast precipitates visible in the BSE micrograph are enriched in Mg and Ce, as seen from respective elemental maps. Color keys with concentration in at.% are provided as insets in the lower left corner of the elemental maps.

Fig. 5 presents the EPMA findings for Mg-3Al-0.5Ce alloy. Unlike Mg-0.5Ce, Mg-3Al-0.5Ce alloy manifests a relatively non-uniform distribution of stringers-like precipitates aligned along the extrusion direction. These precipitates are of varying size, ranging from sub-micron to tens of microns, and exhibit different compositions, including complex Mg-Al-Ce and $Al_{11}Ce_3$ phases. Furthermore, Al is also present in the matrix, seen as clouds around the stringer precipitates in Fig. 5d. The formation of Al-Ce-rich precipitates instead of Mg-Ce ($Mg_{12}Ce$), as observed in Mg-0.5Ce, confirms Ce's higher affinity for Al over Mg. Therefore, as for Mg-3Al-0.5Ce, the precipitates are relatively coarser and not uniformly distributed in comparison to Mg-0.5Ce alloy, the weakening of basal texture and reduction in grain size is less pronounced in the former than in the latter (Fig. 1b and Fig. 3d). It is noteworthy that the decrease in basal texture and grain size is majorly observed in regions close to the precipitates. Altogether, it can be suggested that Al's presence in the Mg matrix hampers the positive effects of Ce alloying, such as reducing basal texture and grain size. Similar observations have also been reported by Luo et al. [27].

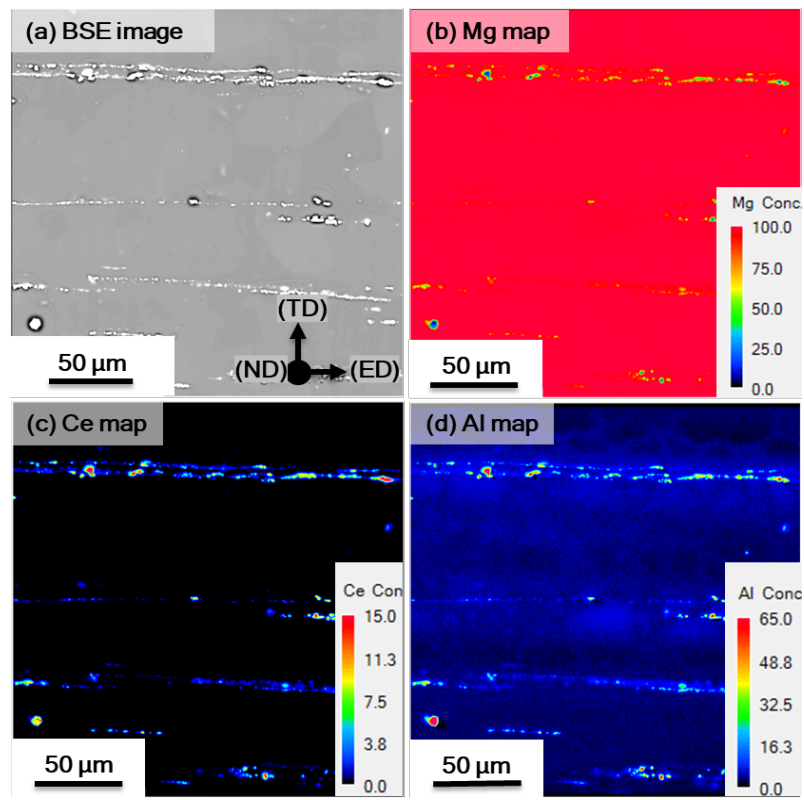

Fig. 5: EPMA results for Mg-3Al-0.5Ce alloy: (a) Backscattered electron (BSE) micrograph, (b) elemental Mg map, (c) elemental Ce map, and (d) elemental Al map. Bright contrast stringer-like precipitates in the BSE micrograph are aligned along the extrusion direction (ED). From elemental maps, these precipitates appear to have complex Mg-Al-Ce and $Al_{11}Ce_3$ compositions. Al clouds around precipitates can also be seen from the Al map in (d). Color keys with concentration in at.% are provided as insets in the lower left corner of the elemental maps.

### 3.2 Effect of Ce addition on the tension-compression behavior

Fig. 6a presents engineering stress-strain curves obtained from tensile testing of Cp-Mg and its investigated Mg-Ce alloys (Mg-0.5Ce and Mg-3Al-0.5Ce) at room temperature (RT) under a strain rate of $10^{-3}$ s$^{-1}$ along the extrusion direction. The investigated Mg-Ce alloys show higher tensile properties than Cp-Mg, as also summarized in Table 1. Specifically, the yield strength of Mg-0.5Ce and Mg-3Al-0.5Ce is 2.15 and 2.24 times higher than that of CP-Mg (~ 64.2 MPa), while their ultimate tensile strength is 1.28 and 1.38 times higher than that of CP-Mg (~ 169.9 MPa). The strain hardening behavior trend is as follows: Cp-Mg > Mg-3Al-0.5Ce > Mg-0.5Ce. Mg-0.5Ce achieves its superior strength over Cp-Mg due to the superimposition of several important strengthening contributions, including solid solution strengthening (due to Ce substitution in the Mg matrix), precipitation hardening (due to the presence of $Mg_{12}Ce$ precipitates) and Hall-Petch grain boundary strengthening (due to change in Hall-Petch constant as a result of Ce segregation at grain boundary and ~ 83% lower grain size than Cp-Mg). Similarly, Mg-3Al-0.5Ce also benefits from these strengthening mechanisms. However, despite having a non-uniform distribution of precipitates and a coarser grain size (~ 50 % larger than Mg-0.5Ce), Mg-3Al-0.5Ce exhibits even higher yield strength than Mg-0.5Ce. This can be attributed to the presence of the Al in the matrix, which provides additional solid solution strengthening [59], as usually reported for conventional Mg-Al alloys [60]. In comparison to previous studies, Mg-3Al-0.5Ce alloy offers similar strength to a hot rolled Mg-Mn-Ce-based ME20 alloy [43] but lower strength than extruded high Gd containing GW103 K alloy with lower grain size [61] and conventional RE-free AZ31 [62] and AM30 [63] alloys. Therefore, to further strengthen present alloys, it is suggested to investigate the additional effects of other favorable elements like Mn, Zn, Y, and Gd.

In addition to their high strength, Mg-0.5Ce, and Mg-3Al-0.5Ce alloys also demonstrate higher ductility (as measured by total elongation-to-failure) than Cp-Mg (Fig. 6a). Table 1 illustrates that the total elongation-to-failure of Mg-0.5Ce and Mg-3Al-0.5Ce is roughly twice that of Cp-Mg. This implies that the addition of Ce can effectively overcome the strength-ductility trade-off in Mg alloys [64]. The improved ductility of Mg-Ce alloys is related to basal texture weakening, which simultaneously activates multiple slip systems, including prismatic and pyramidal [65–67]. This can be correlated with the two scenarios: firstly, there is a reduction of CRSS values on the hard-slipping non-basal systems in Mg alloy, such as the prismatic and pyramidal [68], and second, an increased resistance to dislocation movement on easy-slipping planes, such as basal, which forces the activation of more difficult non-basal slip systems to accommodate plastic deformation [67,68]. The relatively lower ductility of Mg-3Al-0.5Ce compared to Mg-0.5Ce can be ascribed to the limited weakening of the basal texture in the former due to the presence of Al, making it more challenging to activate non-basal slip systems. Moreover, the coarser Al-Ce enriched precipitates introduce strain incompatibility with the matrix at the interface, leading to easy void nucleation and subsequent fracture, as discussed later in relation to fracture surfaces (see Fig. 14). It is worth noting that, despite this, the investigated alloys still exhibit higher ductility compared to extruded Gd-based GW103 K [61], AZ31 [62] and AM30 [63] alloys, suggesting a superior effect of Ce addition.

To determine the presence of asymmetry in Cp-Mg and its alloys, quasi-static compression tests were carried out. The resulting compression engineering stress-strain curves in Fig. 6b illustrate a typical S shape with extensive work hardening in the final stage. The investigated Mg-Ce alloys exhibit higher compression yield strength (CYS) than Cp-Mg. However, as shown in Table 1, the CYS of Cp-Mg and Mg-3Al-0.5Ce is lower than their respective tensile yield strength (TYS), revealing a tension-compression yield asymmetry. Conversely, Mg-0.5Ce displays strikingly similar CYS and TYS values. This trend becomes more evident upon plotting engineering stress-strain curves obtained under tension and compression loading for all three investigated materials together, see Fig. 7. The yield strength ratio (CYS/TYS) was calculated to evaluate the extent of tensile-compression asymmetry, with a ratio closer to 1 suggesting no asymmetry. Upon evaluation, the yield strength ratios for Cp-Mg, Mg-3Al-0.5Ce, and Mg-0.5Ce are 0.56, 0.78, and 1.01, respectively. This indicates that the pronounced tension-compression asymmetry evident for Cp-Mg is reduced in Mg-3Al-0.5Ce and eliminated in Mg-0.5Ce.

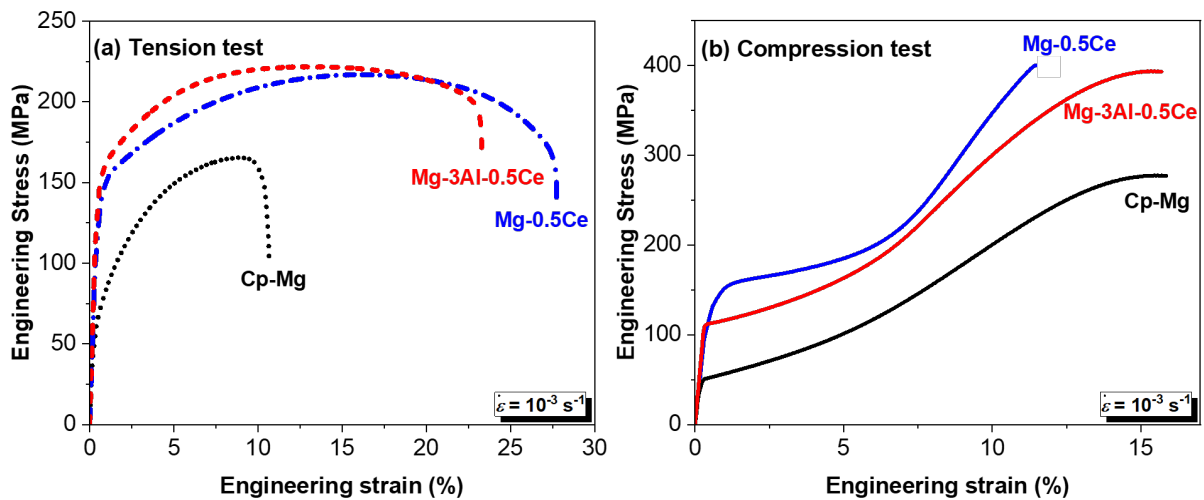

Fig. 6: Representative engineering stress-strain curves obtained upon (a) tensile testing and (b) compression testing Cp-Mg and its alloys, Mg-0.5Ce and Mg-3Al-0.5Ce, under a nominal strain rate of $10^{-3}$ s$^{-1}$. Mg-Ce alloys show improved tensile and compression response than Cp-Mg.

Table 1: Mechanical properties of Cp-Mg, Mg-0.5Ce, and Mg-3Al-0.5Ce under uniaxial monotonic loading.

| Material | Tensile yield strength (MPa) | Ultimate tensile strength (MPa) | Tensile total elongation to failure (%) | Compression yield strength (MPa) |
|---|---|---|---|---|
| Cp-Mg | 64.2 ± 8.2 | 169.9 ± 13.1 | 10.7 ± 2.6 | 35.7 ± 1.1 |
| Mg-0.5Ce | 137.8 ± 2.6 | 218.1 ± 0.9 | 27.8 ± 0.5 | 138.6 ± 4.4 |
| Mg-3Al-0.5Ce | 143.6 ± 12.4 | 233.8 ± 11.6 | 23.3 ± 2.5 | 112.2 ± 7.2 |

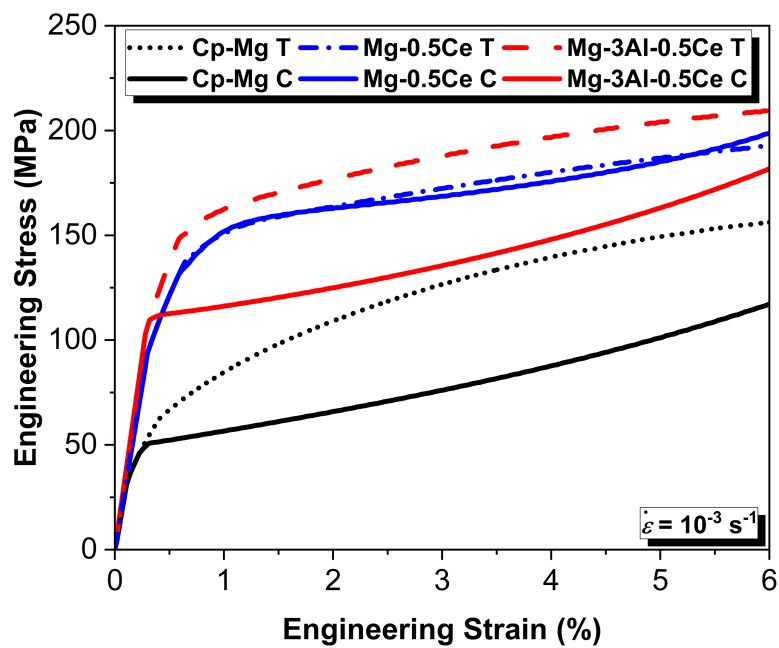

Fig. 7: Representative engineering stress-strain curves obtained under tension (dotted, dotted-dash, and dash lines) and compression (solid lines) loading show tensile-compression yield asymmetry in Cp-Mg, which is comparatively reduced in Mg-3Al-0.5Ce and almost eliminated in Mg-0.5Ce.

To reveal the fundamental reasons behind the investigated materials' monotonic response, post-deformation microstructural investigations were carried out. Fig. 8a and Fig. 8b show the representative IPF maps acquired via EBSD for Cp-Mg post-tensile and post-compression deformations, respectively. At room temperature, Cp-Mg under tensile loading deforms mainly by slip. This is evident in the form of localized misorientation within each grain, manifesting as variations in color among individual grains, as illustrated in Fig. 8a. Furthermore, Fig. 9 shows a near basal oriented grain featuring distinct slip traces. Based on the slip trace analysis described in Section 2.2. and Table 2, the visible slip traces were determined to be from one of the basal slip systems. Similarly, slip-assisted deformation in the other examined grains was found to be dominantly basal. This tendency can be attributed to the lower CRSS value associated with the basal slip systems of Cp-Mg when compared to the non-basal prismatic and pyramidal systems [5–7].

In addition to slip, several twins (with ~ 0.09 area fraction) are observed close to the grain boundaries in Cp-Mg, see Fig. 8a. The misorientation profile across these twins suggests them to be extension twins, specifically $\{10\bar{1}2\}$ $<10\bar{1}1>$ twins, characterized by typical 86° misorientation with respect to the parent grain, see Fig. 8c. The formation of extension twins at first appears counterintuitive, given the context of tensile loading along the extrusion direction. In this scenario, most of the basal-oriented grains, having their c-axes oriented perpendicular to extrusion/loading direction, would conventionally lead to the formation of contraction twins, specifically $\{10\bar{1}1\}$ $<10\bar{1}2>$, with typical 56° misorientation with the parent grain. However, a different scenario could emerge when considering the impact of tension-induced slip deformation and the potential rotation of specific grains within Cp-Mg. This action might induce a compressive strain directed perpendicular to the c-axes of the adjacent grains, consequently setting off the initiation of extension twins in these crystals. Moreover, it's noteworthy that the critical stress threshold needed to activate contraction twinning (~ 85 MPa for Mg's single crystal [69]) is significantly higher compared to the activation threshold for extension twins (~ 7 MPa for Mg's single crystal [69]). This distinction in critical stress levels implies that, given the lower tensile yield strength of Cp-Mg (see Table 1), the circumstances are more conducive for initiating extension twins rather than the contraction twins. Furthermore, Cp-Mg also exhibits the highest strain hardening (~ 164.6%), followed by Mg-3Al-0.5Ce (~ 62.8%) and Mg-0.5Ce (~ 58.3%), which can be attributed to the concurrent activation of slip and twinning mechanisms in Cp-Mg, coupled with its largest grain size that facilitated accommodation of large deformation.

Under compressive loading, Cp-Mg demonstrates slip deformation combined with a noticeably higher fraction of extension twinning, in comparison to under tensile loading (Fig. 8b). Notably, the area fraction of extension twins' (~ 0.17) formed under compression is significantly higher (~ 89.0%) than that observed under tension (compare Fig. 8a and Fig. 8b). The typical S-shaped curves observed in the compression tests can be attributed to the dominant extension twinning phenomena. As twinning results in a fragmentation of the initial grains, the dynamic reduction in grain size contributes to the observed work hardening in the later stage. The formation of extension twins under compression loading is as expected, which can be attributed to the extension of the c-axis when magnesium's HCP lattice is subjected to compressive strain along the basal planes (lying along the extrusion direction). Furthermore, since extension twinning can be activated at comparatively lower stresses and is relatively pronounced under compression, the yield strength of Cp-Mg under compression is lower than that under tension, resulting in tension-compression yield asymmetry.

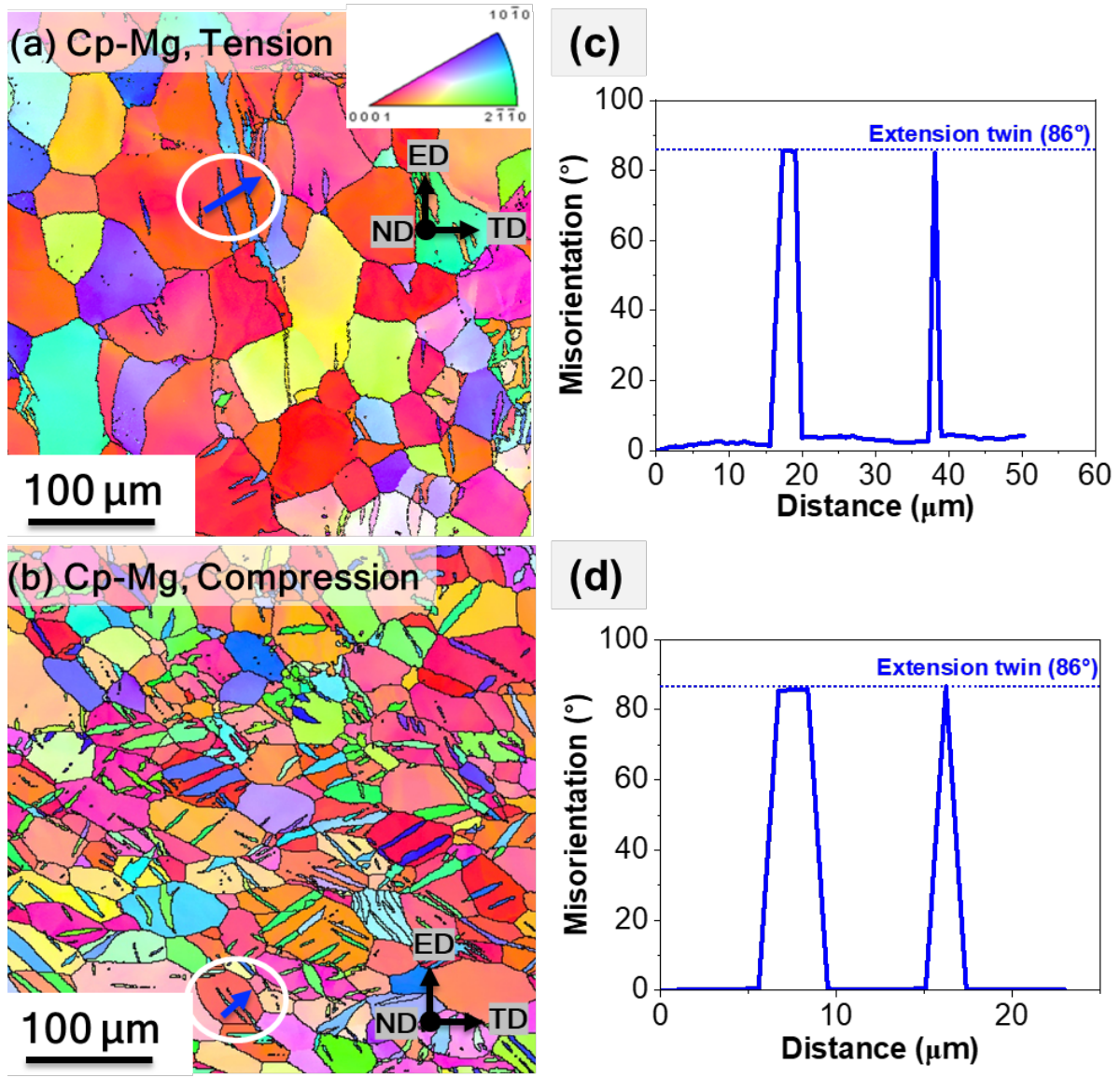

Fig. 8: Representative IPF maps acquired via EBSD, (a) post-tensile and (b) post-compression deformations of Cp-Mg specimens along the extrusion direction. The variation in orientation observed inside each grain suggests slip activation. The twins visible in the IPF maps are found to be extension twins. For instance, the misorientation profiles in (c) and (d) taken across twins marked in (a) and (b), respectively, confirm typical 86° misorientation, suggesting them to be extension twins. Twinning is significantly higher under compression than under tension.

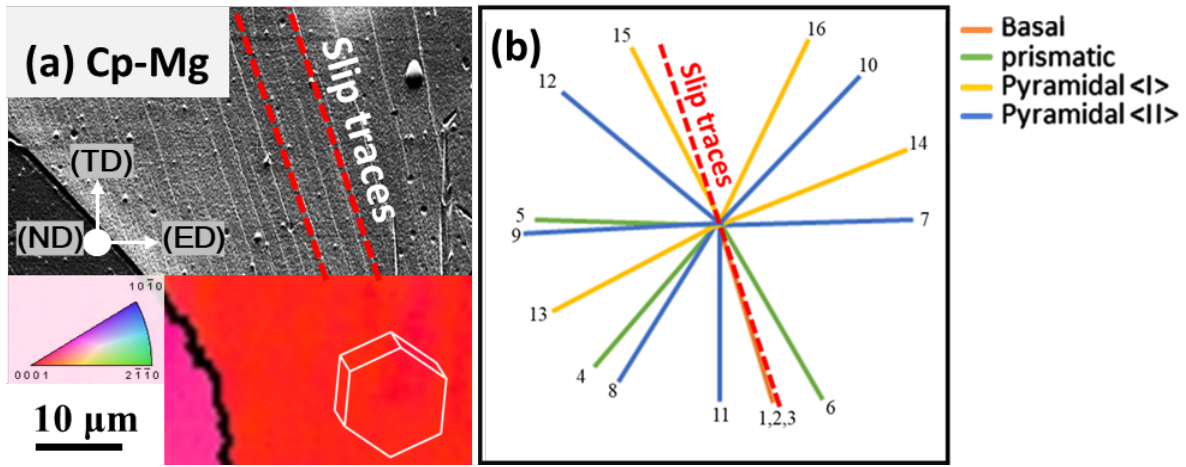

Fig. 9: Illustration of the slip trace analysis performed on a grain from the tensile tested Cp-Mg specimen up to 3% strain. (a) SEM micrograph and overlaid IPF map reveal slip traces inside near basal oriented grain with Euler angles (17.2°, 40.5°, 10.1°). All possible slip traces are plotted in slip system chart (b) based on the orientation and loading direction. Subsequently, note that basal slip system 2 was the selected slip system based on the match, corresponding to (0001) [$\bar{1}2\bar{1}0$]. The Schmid factors for the slip systems are summarized in Table 2. Note that because the slip traces for the slip systems numbered 1, 2, and 3 are parallel to each other, the one (slip system 2) with the highest Schmid factor was chosen for basal slip in the analysis.

Table 2: Schmid factors for the possible slip systems of the analyzed grains of Cp-Mg, Mg-3Al-0.5Ce, and Mg-0.5Ce tensile tested specimens up to 3% strain using dedicated MATLAB code [50].

| Type | Slip traces | Slip system | Schmid Factor | | | |
|---|---|---|---|---|---|---|
| | | | Cp-Mg | Mg-3Al-0.5Ce | Mg-0.5Ce (Grain 1) | Mg-0.5Ce (Grain 2) |
| Basal < a > | 1 | (0001) [11$\bar{2}$0] | 0.26 | 0.28 | 0.19 | 0.19 |
| | 2 | (0001) [$\bar{1}2\bar{1}0$] | 0.49 | 0.20 | 0.26 | 0.10 |
| | 3 | (0001) [2$\bar{1}\bar{1}$0] | 0.23 | 0.48 | 0.45 | 0.08 |
| Prismatic < a > | 4 | (01$\bar{1}$0) [2$\bar{1}\bar{1}$0] | 0.25 | 0.06 | 0.06 | 0.37 |
| | 5 | (10$\bar{1}$0) [$\bar{1}2\bar{1}0$] | 0.02 | 0.24 | 0.33 | 0.45 |
| | 6 | ($\bar{1}$100) [$\bar{1}\bar{1}$20] | 0.28 | 0.30 | 0.27 | 0.08 |
| Pyramidal <c+a> | 7 | (11$\bar{2}$2) [11$\bar{2}\bar{3}$] | 0.02 | 0.20 | 0.16 | 0.33 |
|  | 8 | ($\bar{1}$2$\bar{1}$2) [$\bar{1}$2$\bar{1}\bar{3}$] | 0.32 | 0.03 | 0.09 | 0.08 |
|  | 9 | ($\bar{2}$112) [$\bar{2}$11$\bar{3}$] | 0.01 | 0.33 | 0.39 | 0.10 |
|  | 10 | ($\bar{1}\bar{1}$22) [$\bar{1}\bar{1}$2$\bar{3}$] | 0.21 | 0.05 | 0.01 | 0.49 |
|  | 11 | (1$\bar{2}$12) [1$\bar{2}$1$\bar{3}$] | 0.12 | 0.21 | 0.14 | 0.17 |
|  | 12 | (2$\bar{1}\bar{1}$2) [2$\bar{1}\bar{1}\bar{3}$] | 0.21 | 0.10 | 0.02 | 0.03 |
| Pyramidal <a> | 13 | (01$\bar{1}$1) [2$\bar{1}\bar{1}$0] | 0.12 | 0.28 | 0.16 | 0.36 |
|  | 14 | ($\bar{1}$011) [$\bar{1}$2$\bar{1}$0] | 0.25 | 0.11 | 0.17 | 0.35 |
|  | 15 | (1$\bar{1}$01) [$\bar{1}\bar{1}$20] | 0.37 | 0.39 | 0.33 | 0.02 |
|  | 16 | (0$\bar{1}$11) [2$\bar{1}\bar{1}$0] | 0.33 | 0.17 | 0.27 | 0.29 |

Mg-3Al-0.5Ce alloy exhibits similar deformation behavior as Cp-Mg. During tensile loading, slip is accompanied by twinning, as depicted in Fig. 10a. Fig. 11 displays a deformed grain revealing distinct slip traces. Through slip trace analysis, it was determined that the visible slip traces belong to one of the basal slip systems. Like Cp-Mg, slip-assisted deformation in the other examined grains of Mg-3Al-0.5Ce was also found to be dominantly basal. The restricted activation of non-basal slip systems in Mg-3Al-0.5Ce can be attributed to the constrained weakening of the basal texture. This primarily is due to the presence of Al, which with Ce forms Al-Ce enriched precipitates. As a result, Ce becomes unavailable in the matrix for lowering the CRSS value of non-basal slip systems; and thus, making them more challenging to be activated. Apart from dominant basal slip, the twins' area fraction after tensile loading was determined to be ~ 0.12. Twins were found to be of both extension and contraction type (see Fig. 10a, c). Notably, two types of contraction twins: (i) {10$\bar{1}$1} <10$\bar{1}$2>, with typical 56° misorientation with the parent grain and (ii) {10$\bar{1}$3} <10$\bar{1}$2>, with typical 64° misorientation with the parent grain were observed. The presence of extension twins in this alloy can be attributed to the same underlying reason mentioned earlier for Cp-Mg. Additionally, the occurrence of contraction twins is linked to its higher tensile yield strength (~ 143.6 MPa), as evidenced in Fig. 6a and Table 1, which raises the stress in the localized region to the critical twinning stress required for the formation of contraction twins. Moreover, using molecular dynamics simulations, Tang et al. [70] showed a decrease in the stress required for twin formation in magnesium with Al addition (up to 3wt.%), which further justifies the presence of contraction twins and an overall higher fraction of twins than that observed in Cp-Mg. Furthermore, the presence of Al is also suggested to lower the stacking fault energy (SFE) of magnesium [71,72]. Nevertheless, whether the decrease in Mg's SFE influences the twinning phenomenon remains unclear.

When subjected to compression, Mg-3Al-0.5Ce accommodates deformation through slip along with significant extension twinning. The area fraction occupied by twins under compression is ~ 0.19, around ~ 58.3% higher than the combined fraction of contraction and extension twins observed under tension. And since extension twinning can be activated at relatively low stresses and is fairly pronounced under compression, the compression yield stress is lower than the tensile yield strength for Mg-3Al-0.5Ce, resulting in tension-compression yield asymmetry [73–75]. However, compared to Cp-Mg, the difference between the area fraction of twinning in two loading directions is ~ 12.5% lower for Mg-3Al-0.5Ce. This leads to a slight reduction in tension-compression yield asymmetry in Mg-3Al-0.5Ce compared to Cp-Mg.

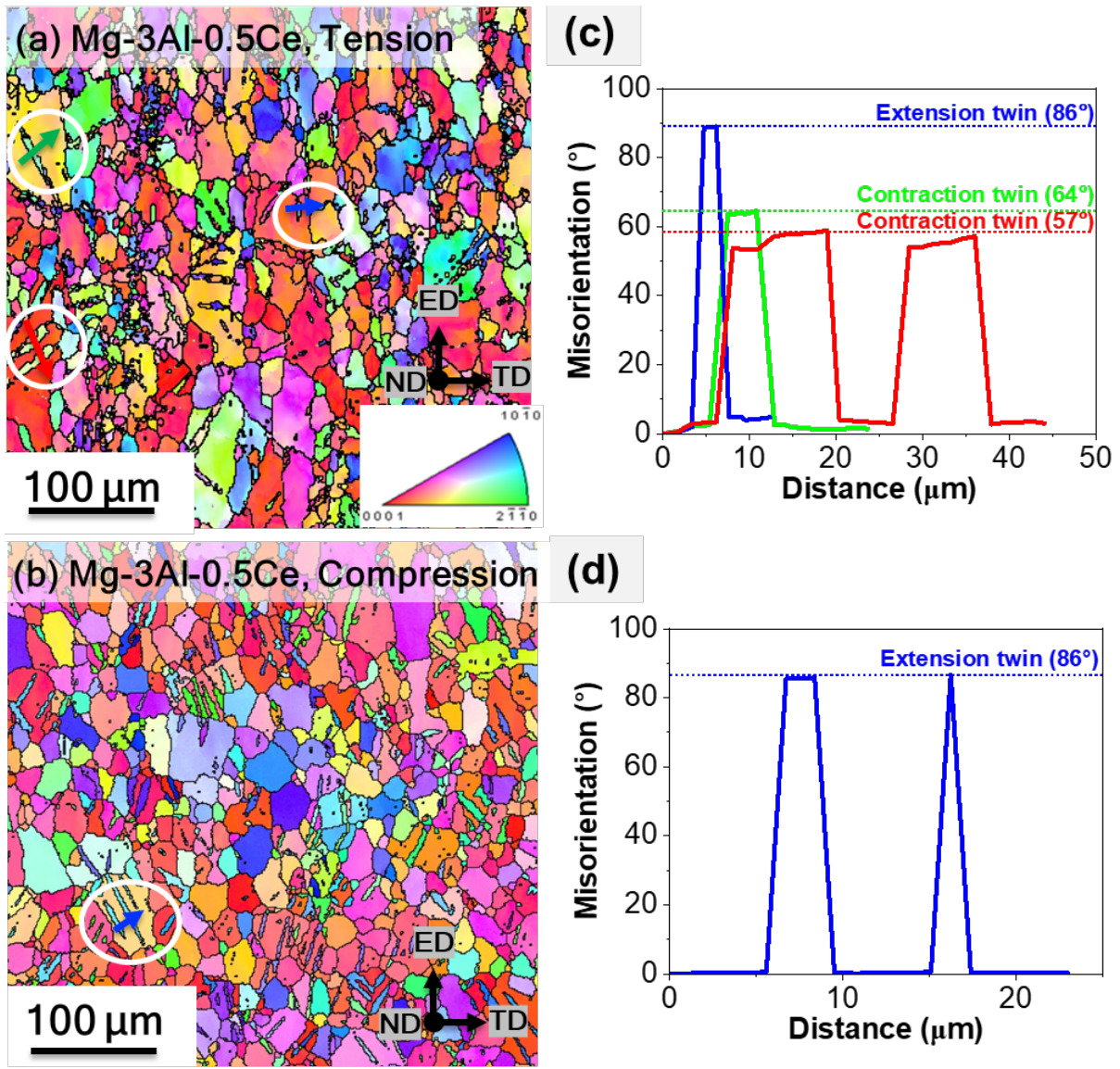

Fig. 10: Representative IPF maps acquired via EBSD, (a) post-tensile and (b) post-compression deformations of Mg-3Al-0.5Ce specimens along the extrusion direction. The variation in orientation observed inside each grain suggests slip activation. Twins' visible post-tensile deformation in (a) are found to be both extension and contraction twins. The misorientation profiles in (c) taken across marked twins in (a) confirm typical 86° misorientation of the marked extension twin and 64° and 56° misorientations for the marked contraction twins. All twins' visible post-compression deformation in (b) are found to be extension twins. Misorientation profiles in (d) taken across marked twins in (b) confirm typical 86° misorientation of the extension twins. Twinning is relatively higher under compression than under tension.

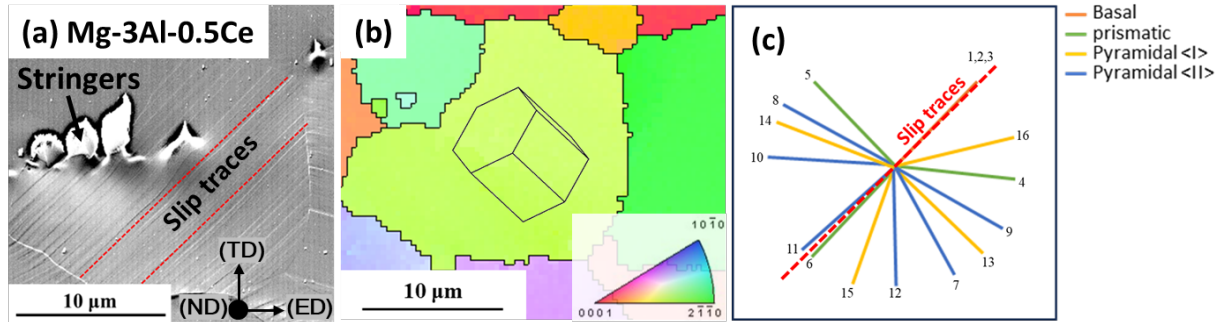

Fig. 11: Illustration of the slip trace analysis performed on grain from the tensile tested Mg-3Al-0.5Ce specimen up to 3% strain. SEM micrograph (a) along with IPF map (b) reveals slip traces inside a grain with Euler angles (136.5°, 57.2°, 24.8°). All possible slip traces are plotted in slip system chart (c) based on the orientation and loading direction. Subsequently, note that basal slip system 3 was selected based on the match, corresponding to (0001) [$2\bar{1}\bar{1}0$]. The Schmid factors for the slip systems are summarized in Table 2. Note that because the slip traces for the slip systems numbered 1, 2, and 3 are parallel to each other, the one (slip system 3) with the highest Schmid factor was chosen for basal slip in the analysis.

In contrast to Cp-Mg and Mg-3Al-0.5Ce, the deformation behavior of Mg-0.5Ce is primarily dominated by slip, with a minor contribution from twinning under both tension and compression loading. This is evident from the IPF maps acquired from the specimens tested under tension and compression loading, as depicted in Fig. 12a and Fig. 12b, respectively. Furthermore, Fig. 13 displays two grains labeled 1 and 2, each exhibiting distinct slip traces. Through slip trace analysis, it was determined that the visible slip traces in grain 1 belong to one of the basal slip systems, while those in grain 2 correspond to one of the pyramidal slip systems. In contrast to the Cp-Mg and Mg-3Al-0.5Ce, among the deformed grains examined for Mg-0.5Ce, the observed slip traces were both from the basal and non-basal slip systems. This can be attributed to the significant weakening of the basal texture in Mg-0.5Ce, as shown in Fig. 1b and Fig. 3g, which appears to favorably lower the CRSS value of activating non-basal slip systems (including prismatic and pyramidal). Thereby resulting in the activation of multiple slip systems. This is also linked to the fact that the formation of extension twins in Mg-0.5Ce is only sporadic and not prominent under both tensile and compression loading (Fig. 12). The suppressed twinning in Mg-0.5Ce can be attributed to the Ce effect, which may increase the CRSS value required for twinning activation and/or lower grain size. It is well known that the finer the grain size, the higher the stress needed for activating twinning, and vice versa. Consequently, plastic deformation in Mg-0.5Ce is mainly accommodated by slip at multiple-slip systems, leading to similar yield strengths under both tension and compression loading, thereby eliminating tension-compression yield asymmetry. Furthermore, multiple-slip systems activation also assists Mg-0.5Ce in accommodating large plastic strain before fracture, resulting in highest tensile ductility among investigated Cp-Mg and Mg-3Al-0.5Ce.

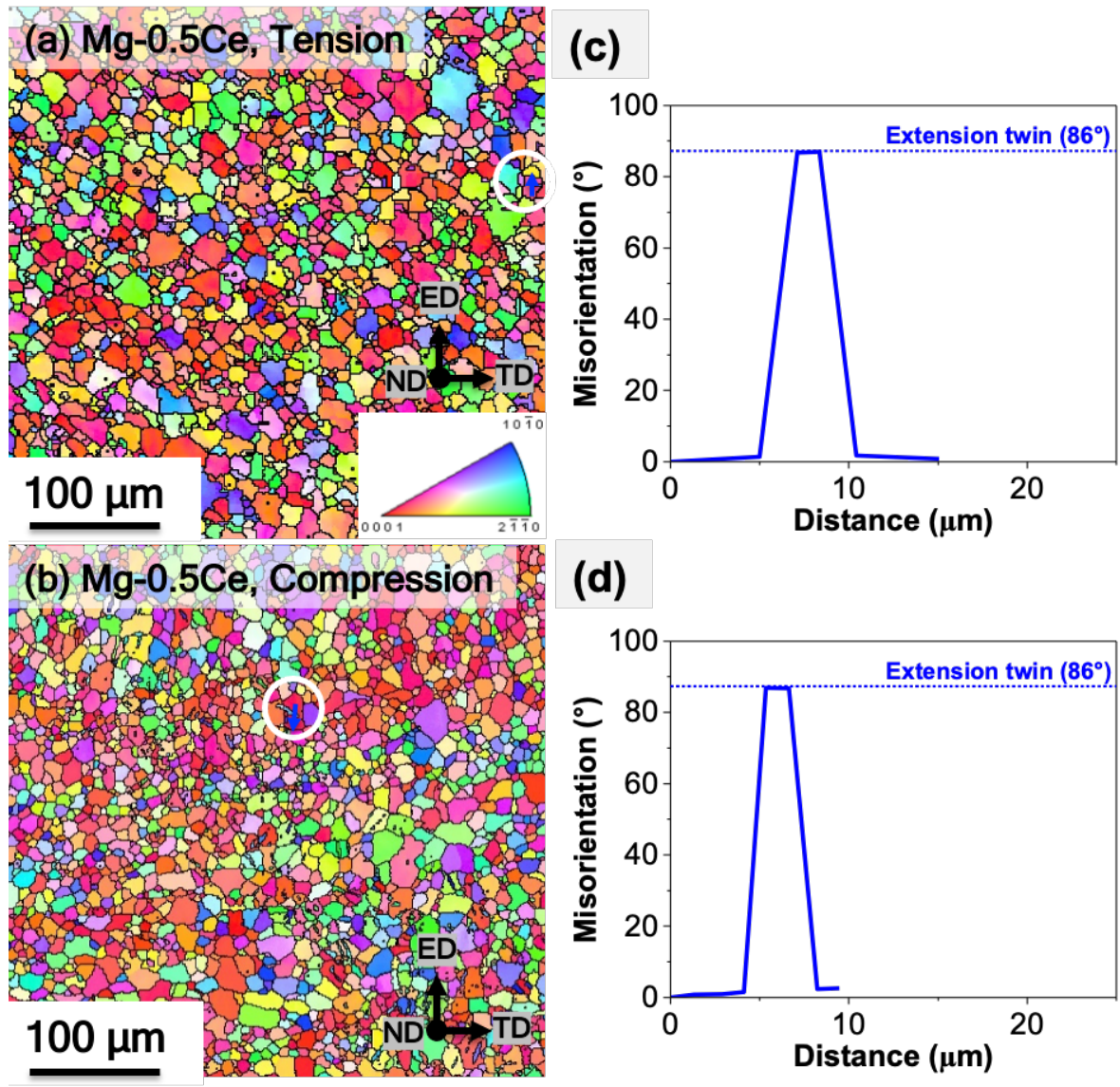

Fig. 12: Representative IPF maps acquired via EBSD (a) post-tensile and (b) post-compression deformations of Mg-0.5Ce specimens along the extrusion direction. The variation in orientation observed inside each grain suggests that slip controls deformation. The sporadic twins visible in the IPF maps are found to be extension twins. For instance, the misorientation profiles in (c) and (d) taken across twins marked in (a) and (b), respectively, confirm typical 86° misorientation, suggesting them to be the extension twins.

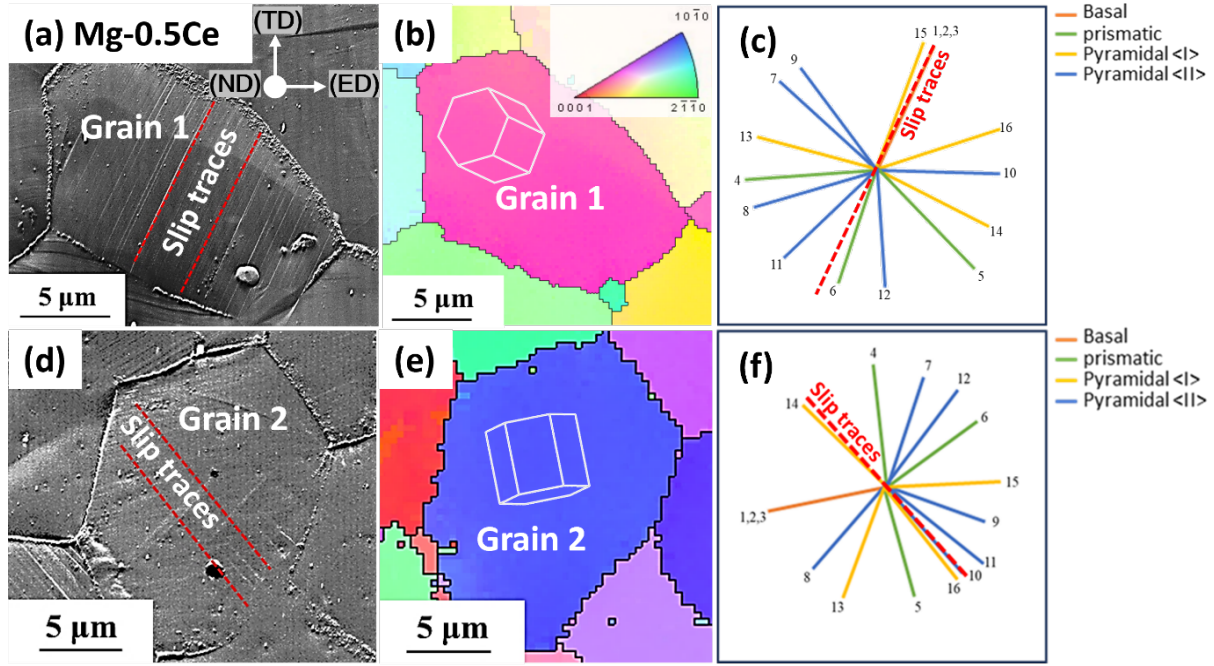

Fig. 13: Illustration of the slip trace analysis performed on two grains from the tensile tested Mg-0.5Ce specimen up to 3% strain. SEM micrographs (a, d) along with IPF maps (b, e) revealing slip traces inside grains 1 and 2 with Euler angles (155°, 36.3°, 5.0°) and (101°, 97.8°, 356.7°), respectively. All possible slip traces are plotted in slip system charts (c, f) based on the orientations and loading direction. Subsequently, note that for grain 1, basal slip system 3 was selected based on the match, corresponding to (0001) [$2\bar{1}\bar{1}0$]. Similarly, for grain 2, pyramidal slip system 10 was selected based on the match, corresponding to ($\bar{1}\bar{1}22$) [$\bar{1}\bar{1}2\bar{3}$]. The Schmid factors for the slip systems are summarized in Table 2. Note that because the slip traces for the slip systems numbered 1, 2, and 3 are parallel to each other, the one (slip system 3) with the highest Schmid factor was chosen for basal slip in the analysis.

Fig. 14 displays SEM micrographs of the fracture surfaces from fractured tensile specimens of Cp-Mg and its Mg-Ce alloys (Mg-0.5Ce and Mg-3Al-0.5Ce). As shown in Fig. 14a, the fracture surface of the Cp-Mg tensile specimen exhibits a lamellar faceted morphology, indicative of a cleavage-like brittle fracture associated with its poor ductility. Conversely, the SEM micrographs captured from the fractured tensile specimens of both Mg-0.5Ce and Mg-3Al-0.5Ce alloys in Fig. 14b and Fig. 14c reveal rough fracture surfaces with a high density of dimples. This observation suggests that the fracture in Mg-Ce alloys is primarily preceded by extensive localized plastic deformation, correlating enhanced ductility of both Mg-0.5Ce and Mg-3Al-0.5Ce compared to Cp-Mg. The formation of dimples is due to the nucleation and growth of voids in regions of high-stress concentration. Notably, dimples appear to originate from the large stringer-like precipitates in Mg-3Al-0.5Ce, see Fig. 14d, explaining its lower ductility than Mg-0.5Ce.

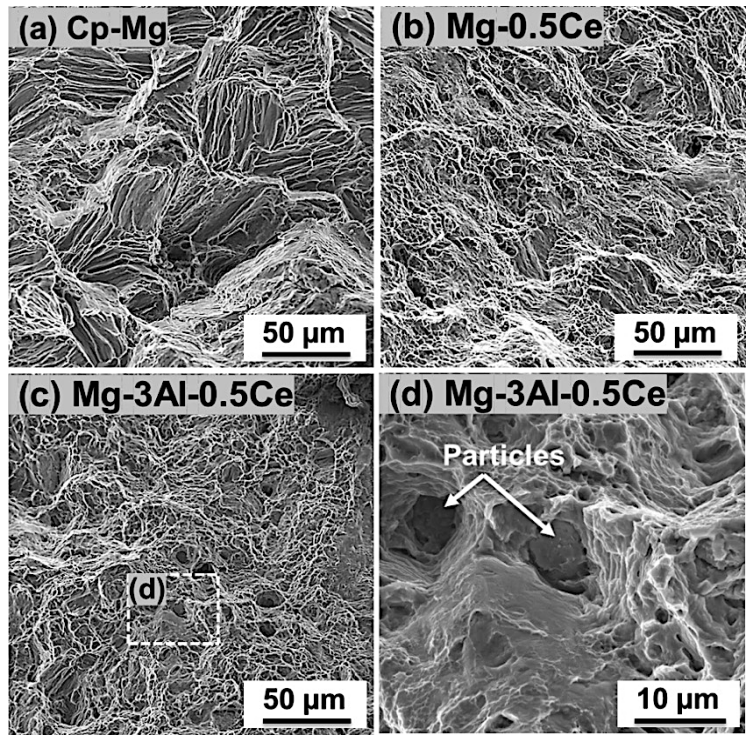

Fig. 14: SEM micrographs revealing fracture surfaces of the fractured tensile specimens of (a) Cp-Mg, (b) Mg-0.5Ce, and (c, d) Mg-3Al-0.5Ce. Lamellar cleavage-like faceted brittle fracture for Cp-Mg and rough fracture surfaces with a high density of dimples are evident for Mg-Ce alloys. (d) A magnified view of the marked region in (c) reveals stringer particles at the center of the dimples from the fractured Mg-3Al-0.5Ce sample.

### 3.3 Effect of Ce addition on the low-cycle fatigue (LCF) response

Fig. 15 shows the representative stress-strain hysteresis loops at second and half-life cycles for LCF-tested Cp-Mg and its investigated Mg-Ce alloys (Mg-0.5Ce and Mg-3Al-0.5Ce) at RT under a strain amplitude of ± 0.5%. It can be observed that the hysteresis loops for Cp-Mg are asymmetric, with significant differences in their tensile and compressive peak stresses at both the second and half-life cycles (752$^{nd}$ cycle). Additionally, the loops exhibit a characteristic sigmoidal and concave-up shape commonly observed in wrought magnesium alloys such as AZ31 and AM30 [76,77], with basal texture.

Both Mg-Ce alloys show higher peak stresses and lower inelastic strain amplitudes than Cp-Mg, as indicated by the reduced width of the hysteresis loops at zero stress. Furthermore, the Mg-3Al-0.5Ce alloy hysteresis loops display lower asymmetry, as the difference between the tensile and compressive peak stresses is significantly lower in the second cycle than that for Cp-Mg, which almost disappears at the half-life (773$^{rd}$ cycle). On the other hand, the Mg-0.5Ce alloy exhibits a nearly symmetrical cyclic response, characterized by similar tensile and compressive peak stresses in the hysteresis loops at both the second and half-life cycles (471$^{st}$ cycle). The cyclic peak stresses at first and half-life cycles, inelastic strain amplitude at half-life cycles, and number of cycles to failure for Cp-Mg, Mg-0.5Ce, and Mg-3Al-0.5Ce are listed in Table 3.

To further analyze the asymmetry, the eccentricity parameter (e) is introduced, defined as the horizontal distance between the intersection point of a line connecting the tensile and compressive peaks of the stress-strain hysteresis loop and the coordinate origin, as shown in Fig. 15. The eccentricity values for Cp-Mg, Mg-0.5Ce, and Mg-3Al-0.5Ce are assigned $e_M$, $e_C$, and $e_A$, respectively. Notably, perfectly symmetrical hysteresis loops, such as those observed in face-centered cubic (FCC) metals like copper, nickel, and aluminum, would have eccentricity values close to zero. From Fig. 15, it can be observed that the highly asymmetric and skewed hysteresis loops of Cp-Mg result in eccentricity values of approximately ~ 0.19% and ~ 0.15% for the second and half-life (752$^{nd}$) cycles, respectively. In contrast, Mg-0.5Ce displays nominal eccentricity values of ~ 0.015% and ~ 0.004% for the second and half-life (471$^{th}$) cycles, respectively, indicating near symmetrical hysteresis loops. Lastly, the calculated eccentricity values for the Mg-3Al-0.5Ce alloy fall between Cp-Mg and Mg-0.5Ce, with values of ~ 0.05% and ~ 0.01% at the second and half-life (773$^{rd}$) cycles, respectively.

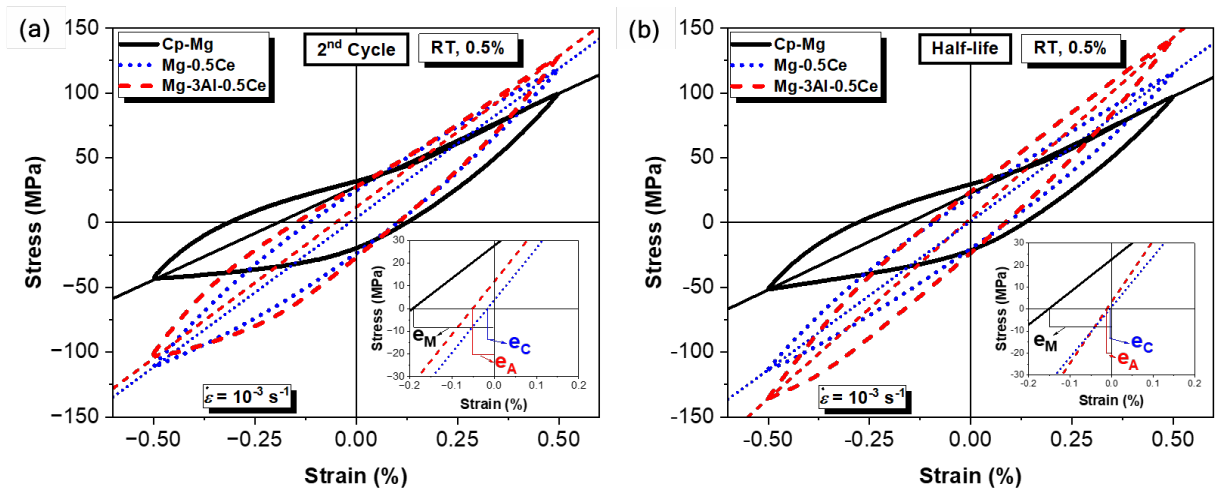

Fig. 15: Hysteresis loops for Cp-Mg, Mg-0.5Ce, and Mg-3Al-0.5Ce specimens tested under ± 0.5% strain amplitude with a strain rate of 10$^{-3}$ s$^{-1}$ at RT at (a) 2$^{nd}$ and (b) half-life cycles. The inset shows the eccentricity parameter (e) evaluated as the horizontal distance between the point of intersection of a line drawn through the tensile and compressive peaks of the stress-strain hysteresis loop and the coordinate origin. The eccentricity values for Cp-Mg, Mg-0.5Ce, and Mg-3Al-0.5Ce are assigned $e_M$, $e_C$, and $e_A$, respectively.

Fig. 16 illustrates the cyclic stress response, including tensile peak, compressive peak, and mean stresses, as a function of the number of cycles for Cp-Mg, Mg-0.5Ce, and Mg-3Al-0.5Ce. It is evident that the addition of Ce increases cyclic peak stresses, as observed from the hysteresis loops. Mg-3Al-0.5Ce exhibits the highest tensile peak stresses, followed by Mg-0.5Ce and Cp-Mg. In the context of the initial compressive peak stresses, Mg-0.5Ce demonstrates the highest values, followed by Mg-3Al-0.5Ce and Cp-Mg. These trends in cyclic stresses align with the previously discussed monotonic mechanical response, as shown in Fig. 6.

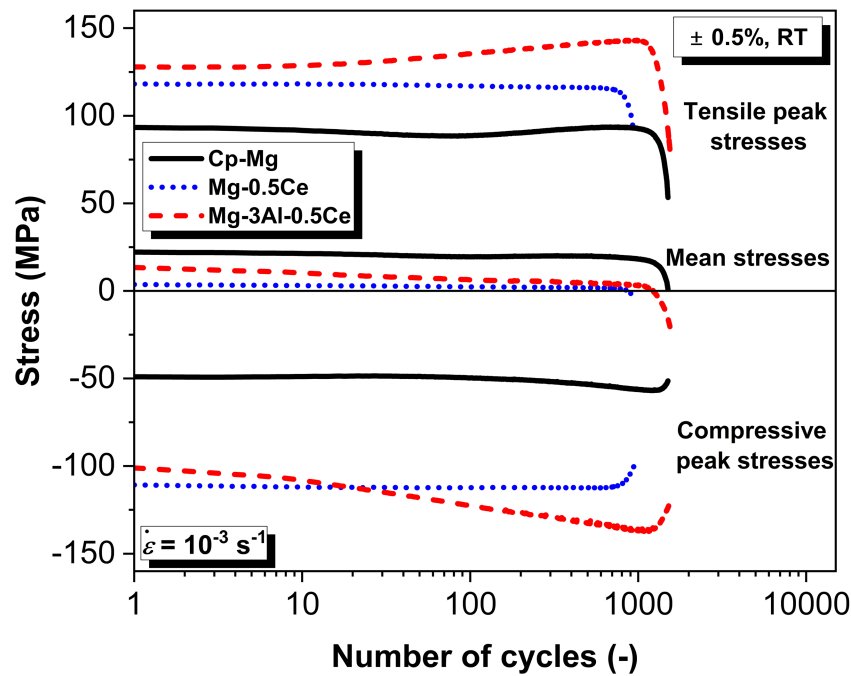

Fig. 16: Representative tensile peak, compressive peak, and mean stresses versus the number of cycles curves for Cp-Mg, Mg-0.5Ce, and Mg-3Al-0.5Ce tested under ± 0.5% strain amplitude with a strain rate of $10^{-3}$ s$^{-1}$ at RT. The mean stresses are calculated as $\sigma_{mean} = (\sigma_{max} + \sigma_{min})/2$. Where $\sigma_{max}$ is the tensile peak stress, and $\sigma_{min}$ is the compressive peak stress.

In the context of the peak stress response, both tensile and compressive peak stresses for Cp-Mg undergo insignificant change initially (i.e., cyclic stable regime) followed by their minor increment (i.e., cyclic hardening stage) until failure. The compressive peak stresses are lower than the tensile peak stresses, resulting in nearly constant positive mean stresses of around 25 MPa, indicating cyclic asymmetry. As discussed in the monotonic testing section, this cyclic asymmetry can also be attributed to the significantly higher area fraction of extension twinning under compression than under tension, leading to lower compression peak stresses. Multiple studies have demonstrated that basal slip is initiated during the initial tensile loading phase, followed by extension twinning in the subsequent compression loading phase [36]. In the second cycle, detwinning occurs alongside basal slip during the tensile loading phase, followed by again extension twinning during the compression loading phase [36]. This microstructural evolution persists, with residual twins progressively cumulated with increasing cycles [36]. Indeed, microstructure characterization carried out after specimen failure revealed extensive residual twins in Cp-Mg (see Fig. 17a). These residual twins are found to be extension twins (see Fig. 17d), leading to the initial grains' fragmentation. As a result, the dynamic reduction in grain size contributes to the observed cyclic hardening in the later stage.

Mg-3Al-0.5Ce also exhibits an initial cyclic stable stress response followed by cyclic hardening until failure. However, Mg-3Al-0.5Ce cyclic stress response reveals few differences compared to Cp-Mg. For instance, compared to compressive peak stresses, the tensile peak stresses are initially higher (~ 26 MPa) but become almost equivalent upon further cycling. This trend can be more easily noted from the mean stress profile, where a drop from 13.4 MPa at the first cycle to 3.83 MPa at the half-life is observed. Additionally, the hardening rate in the compression (0.044 MPa/cycle) is greater than that in the tension (0.019 MPa/cycle), which may explain the decrease in asymmetry with cycling in Mg-3Al-0.5Ce. Comparing post-fatigue Mg-3Al-0.5Ce microstructure (Fig. 17c) to that of Cp-Mg (Fig. 17a) reveals relatively fewer residual extension twins. This observation suggests that a substantial portion of the twins formed during the compression cycle in Mg-3Al-0.5Ce is reversed, or detwinned, during the subsequent tension cycle. This balanced occurrence of twinning and detwinning contributes to the relatively symmetrical cyclic response of Mg-3Al-0.5Ce when compared to Cp-Mg. Moreover, the extension twins that persist after the tension phase tend to increase the hardening effect during the subsequent compression cycle.

Unlike Cp-Mg and Mg-3Al-0.5Ce, the cyclic stress response of Mg-0.5Ce remains relatively stable throughout its fatigue life. At the first and half-life ($471^{st}$) cycles, the tensile peak stresses are 116.0 MPa and 118.1 MPa, respectively. Likewise, the compressive peak stresses are -114.7 MPa and -116.5 MPa, respectively. Notably, the mean stresses are negligible during the entire fatigue life. The comparable tensile and compressive peak stresses for Mg-0.5Ce can be attributed to its initial random texture and finer grain size, as observed in Fig. 1b and Fig. 3d, which results in the activation of similar deformation mechanisms under both loading directions, as discussed in the previous section. For Mg-0.5Ce, deformation is primarily governed by slip mechanisms, both on the basal and non-basal slip systems. Twinning plays a minor role in the overall deformation behavior of the alloy. Fig. 17b indicates that Mg-0.5Ce exhibits the least extent of residual extension twins compared to Cp-Mg and Mg-3Al-0.5Ce. These findings align with previous LCF studies on Mg-RE alloys, such as Mg-10Gd-3Y-0.5Zr (GW103K) and Mg-3Y alloy [29,36]. Besides, as shown in Fig. 16, despite having higher cyclic peak stresses, the Mg-Ce alloys exhibit a comparable fatigue life to Cp-Mg (see Table 3). This can be attributed to the alloys' lower inelastic strain, which plays a crucial role in determining the fatigue life in the LCF regime.

Overall, it can be concluded that 0.5wt.% Ce addition enhances magnesium's cyclic strength, stabilizes its cyclic stress response, lowers inelastic strain, and reduces cyclic asymmetry while maintaining comparable fatigue life.

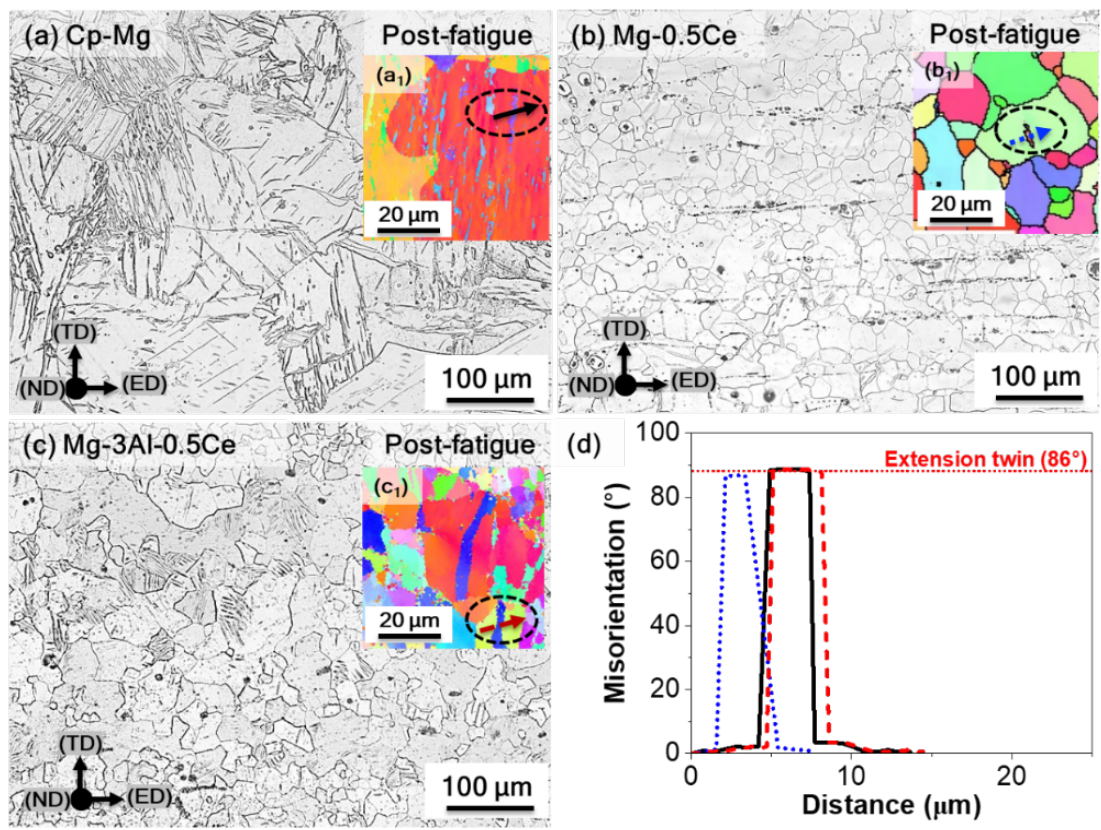

Fig. 17: Optical micrographs revealing representative post-fatigue microstructure acquired from LCF tested (a) Cp-Mg, (b) Mg-0.5Ce, and (c) Mg-3Al-0.5Ce specimens. IPF maps obtained via EBSD from ($a_1$) Cp-Mg, ($b_1$) Mg-0.5Ce, and ($c_1$) Mg3Al0.5Ce specimens along ND are shown as insets. A relatively high density of residual twins within Cp-Mg grains is notably less apparent in Mg-0.5Ce and Mg-3Al-0.5Ce alloy. The twins visible in the IPF maps are found to be extension twins. For instance, the misorientation profiles in (d) taken across twins marked in ($a_1$), ($b_1$), and ($c_1$) insets, respectively, confirm typical 86° misorientation, suggesting them to be extension twins.

Table 3: Cyclic peak stresses at first and half-life cycles, inelastic strain amplitude at half-life cycles, and number of cycles to failure for Cp-Mg, Mg-0.5Ce, and Mg-3Al-0.5Ce under LCF loading at RT under ± 0.5% strain amplitude.

| Material | First cycle | | Half-life cycle | | | Number of cycles to failure |
|---|---|---|---|---|---|---|
| | Peak tensile stress (MPa) | Peak compressive stress (MPa) | Peak tensile stress (MPa) | Peak compressive stress (MPa) | In-elastic strain amplitude (%) | |
| Cp-Mg | 96.6 ± 5.8 | -40.97 ± 6.1 | 96.8 ± 4.7 | -48.8 ± 5.2 | 0.21 ± 0.02 | 1504 ± 110 |
| Mg-0.5Ce | 114.6 ± 2.9 | -110.7 ± 3.7 | 118.1 ± 4.6 | -116.5 ± 3.2 | 0.12 ± 0.01 | 1370 ± 140 |
| Mg-3Al-0.5Ce | 130.9 ± 6.4 | -104.4 ± 4.1 | 138.9 ± 9.3 | -135.0 ± 7.3 | 0.10 ± 0.02 | 1469 ± 109 |

## 4. Summary and conclusions

In this study, we investigate the influence of Ce addition on the microstructure, tension-compression behavior, and cyclic response of pure Mg and Mg-Al alloy at room temperature (RT). The investigated materials Cp-Mg, Mg-0.5Ce, and Mg-3Al-0.5Ce were extruded at 400°C, followed by annealing at the same temperature for one hour. The key findings are as follows:

1. Addition of 0.5wt.% Ce to pure Mg results in weakening its basal texture, formation of uniformly distributed $Mg_{12}Ce$ precipitates, and grain size refinement. As a result, Mg-0.5Ce alloy exhibits improved monotonic properties, such as higher yield strength, ultimate tensile strength, and ductility, compared to Cp-Mg.
2. Addition of 0.5wt.% Ce to pure Mg also leads to the elimination of the tension-compression yield asymmetry. This is associated with the weakening of the basal texture, activation of non-basal slip systems, and the fine grain size, which suppresses twinning activity. Mg-0.5Ce deforms primarily by slip with occasional extension twinning in both tension and compression loading directions.
3. In the presence of 3wt.% Al, the advantages effect of the 0.5wt.% Ce addition in Mg is suppressed. This is due to the inherent Al-Ce affinity, leading to the formation of complex Mg-Al-Ce and $Al_{11}Ce_3$ stringer-like precipitates. The resulting unavailability of Ce in the Mg matrix limits the weakening of the basal texture, grain size reduction, and ductility improvement, as observed for Mg-0.5Ce alloy. Nevertheless, Al addition contributes to the solid solution strengthening and possibly lowers the critical stress required for twinning in Mg, resulting in the highest tensile strength of Mg-3Al-0.5Ce.

4. Like Cp-Mg, Mg-3Al-0.5Ce deforms by slip and twinning, with the contribution of each mechanism dependent on the loading direction. The area fraction of twinning is ~ 58.3% higher under compression than that under tension. Since extension twinning can be activated at relatively low stresses and is fairly pronounced under compression, the compression yield stress is lower than the tensile yield strength for both Mg-3Al-0.5Ce and Cp-Mg, resulting in tension-compression yield asymmetry. However, compared to Cp-Mg, the difference between the area fraction of twinning in two loading directions is ~ 12.5% lower in Mg-3Al-0.5Ce. This leads to a reduction in tension-compression yield asymmetry in Mg-3Al-0.5Ce compared to Cp-Mg.

5. Addition of 0.5wt.% Ce to pure Mg and Mg-Al alloy also enhances their cyclic strength, stabilizes cyclic stress response, reduces inelastic strain, and minimizes cyclic asymmetry while maintaining a comparable fatigue life. A relatively high density of residual extension twins within Cp-Mg grains was notably reduced in Mg-0.5Ce and Mg-3Al-0.5Ce alloys. This microstructural difference is related to the enhanced fatigue response of Mg-Ce alloys.

**Acknowledgments**

AC gratefully acknowledges financial support from the Indian Institute of Science, Bengaluru, and the Science and Engineering Research Board, Department of Science and Technology (DST-SERB, Project code: SERB/F/5812/2022-2023), Government of India. The authors gratefully acknowledge Prof. Satyam Suwas for generously providing extruded bars and access to the XRD and EBSD facilities. The authors also recognize the Advanced Facility for Microscopy and Microanalysis (AFMM) at IISc, Bengaluru, for providing access to the characterization facilities.

**Data availability**

The data supporting this study's findings are available from the corresponding author upon reasonable request.